\documentclass[aps,prl,twocolumn]{revtex4}
\usepackage{graphicx}% Include figure files

\begin{document}

% define terms

%%%%%%%%%%%%%%%%%%%%%%%%%%%%%%%%%%%%%%%%%%%%%%%%%%%%

%\input{macro}

\newcommand{\eqref}[1]{(\ref{#1})}

\newcommand{\be}{\begin{equation}}

\newcommand{\ee}{\end{equation}}

\newcommand{\bea}{\begin{eqnarray}}

\newcommand{\eea}{\end{eqnarray}}

%%%%%%%%%%%%%%%%%%%%%%%%%%%%%%%%%%%%%%%%%%%%%%%%%%%%

%when use two column
%\twocolumn[\hsize\textwidth\columnwidth\hsize\csname@twocolumnfalse\endcsname\draft

\title{Spin filtering in a magnetic barrier structure:\\ in-plane spin orientation}
%ferromagnetic/ferroelectric hybrid double quantum disks}

\author{Nammee Kim\footnote{Electronic mail:
nammee@ssu.ac.kr} and Heesang Kim}
 \affiliation{Department of Physics, Soongsil University, Seoul 156-743, Korea}

\date{\today}

\begin{abstract}
We investigate ballistic spin transport in a two dimensional electron gas system through
 magnetic barriers of various geometries
 using the transfer matrix method.
While most of the previous studies have focused on the effect of magnetic barriers perpendicular to
 the two dimensional electron gas plane,
we concentrate on the case of magnetic barriers parallel to the plane.
We show that resonant oscillation occurs in the transmission probability without
 electrostatic potential modulation which is an essential ingredient in the case of ordinary out-of-plane magnetic barriers.
Transmission probability of the in-plane magnetic barrier structure changes drastically
according to the number of barriers and also according to the electrostatic potential modulation applied in the magnetic barrier region.
Using a hybrid model consisting of a superconductor, ferromagnets, and a two dimensional electron gas plane, we show that it can serve as a good in-plane oriented spin selector which can be operated thoroughly by electrical modulation without any magnetic control.
\end{abstract}

%\\
%Key Words: Magnetic Semiconductor, Hybrid Quantum Structure,
%Ferroelectric Semiconductor, Double Quantum Well \\
%PACS number: 75.50.Pp, 75.75.+a}
%\pacs{75.50.Pp, 75.75.+a}

%%%%%%%%%%%%%%%%%%%

\maketitle

%]

% "]" should be commented with twocol....

Semiconductor device including magnetic barriers has recently attracted much attention
as a spin device, because it circumvents the resistance mismatch problem in the spin injection process, which is one of the main
obstacles in realization of the Datta-Das-type spin transistor\cite{datta}.
Very recently the magnetic barrier study is expanded to a graphene\cite{graphene} and a topological insulator\cite{topo} with great interests.
Magnetic barrier structure has been introduced by using vortices in superconductors, superconducting(SC) masks or ferromagnetic material stripes on
a two dimensional electron gas(2DEG).
Since the 2DEG having a perpendicular magnetic field has been studied intensively in experiments,
for example, Quantum Hall effect,  observation of Commensurability effects and  Novel giant
 magnetoresistance,
 most of the previous theoretical studies\cite{peeters,ihm, nammee1}
 on a magnetic barrier structure have carried on the out-of-plane magnetic barrier system
 with a purpose to use the system as an out-of plane oriented spin filter or a spin injection device.
% even though it has a crucial requirement of electrostatic modulation
%to achieve a satisfiable spin polarization effect.
% by neglecting  possible in-plane magnetic field component.
%
However, in some experiments like those on Spin valve and Spin Hall Effect\cite{kimura,tinkham,Lu,Wees},
spin orientation is along the 2DEG plane and an in-plane oriented spin filter or
a spin injection device is in order, i.e., an in-plane magnetic barrier system start to draw attention.\cite{bijl, zhai,wang}.

The aim of our work is to investigate the ballistic transport properties
through an in-plane magnetic barrier system in 2DEG.
In this system, it is easier to have tunneling process compared to an out-of-plane magnetic barrier system
because the in-plane barrier system has no unwanted magnetic barrier due to vector potential\cite{ihm}.
We calculate spin-dependent transmission coefficients for double and triple
in-plane magnetic barrier systems
with/without external electrostatic modulation across a barrier using transfer matrix method.
Schematic illustration for the possible realization of the in-plane magnetic barrier system
 is shown in Fig. 1(a).
The device consists of  ferromagnets(FM) and a superconducting(SC) mask having openings on top of
a 2DEG system. The SC mask is used to provide the magnetic field profile demanded in the
 system as shown in Fig. 1(b).
 \begin{figure}
\includegraphics[width=8.0cm]{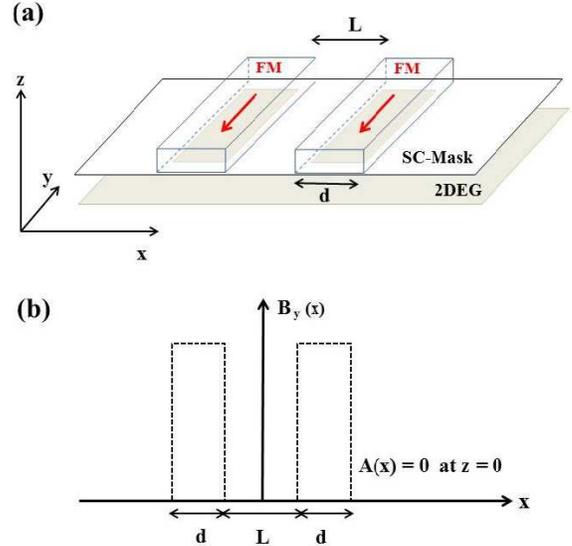}
\noindent \caption{(a) Schematic illustration for a possible realization of the in-plane magnetic field barriers.
The device consists of  ferromagnets and a SC mask  on top of
 a 2DEG system. $L$ is the distance between openings in the SC mask  and $d$ is the opening width.  (b) Dotted line indicates the in-plane  magnetic field profile in the 2DEG under the SC mask. }
\label{Figure1}
\end{figure}

The Hamiltonian with effective mass $m^*$, and effective $g$-factor $g^*$ with electrostatic potential $U(x)$ is;
\be
H= \frac{(\vec{p}+ e/c \vec{A})^2}{2m^*}+U(x) + \frac{e\hbar g^* \sigma}{4m_{0}c}B_{y}(x),
\label{hamiltonian}
\ee
where $\sigma=+1/-1$ denotes spin up/spin down of the electron.
The in-plane magnetic barrier,
which is  square-function like, assumes the magnetic field $B_{y}$ along the y direction at two locations $x=-L/2$ and $L/2$;
\be
B_{y}(x)=[B\Theta(|x|-L/2) \Theta(L/2+d-|x|)].
\label{b}
\ee
Here, $\Theta(x)$ is the Heaviside step function, B is the magnetic field strength in barriers,
$L$ is the distance between the openings in the SC mask  and $d$ is the width of each opening.
In Fig. 1(b), the dotted line indicates the in-plane magnetic profile in the 2DEG under
the SC mask as written in Eq.\eqref{b}.
Vector potential $\vec{A}$, in Landau gauge, is given by
%\bea
%\vec{A}&=&\int_{0}^{z} B_{y}(x) dz^\prime \,\hat{i}\label{vecpot} \\
%&=&[B\Theta(|x|-L/2) \Theta(L/2+d-|x|)]z \,\hat{i}, \nonumber
%\eea
%
$\vec{A}=\int_{0}^{z} B_{y}(x) dz^\prime \,\hat{i}\label{vecpot}
=[B\Theta(|x|-L/2) \Theta(L/2+d-|x|)]z \,\hat{i}$,
which vanishes at the 2DEG plane ($z=0$). As a result, the transverse motion is decoupled from the longitudinal one.
This formalism also applies to $\vec{B}=B(x)\hat{i}$ with $\vec{A}=\int_{0}^{z} B_{x} dz^{\prime}\,\hat{j}$, which provides longitudinal spin orientation  in 2DEG.

The system is translation-invariant along the y direction, and the Schr\"odinger equation $H\Psi(x,y)=E\Psi(x,y)$
 in two dimensional space is simplified by using $\Psi(x,y)=e^{i k_{y} y}\psi(x)$;
\be
[\frac{d^2}{dx^2}-k_{y}^2 +2(E-U(x))-\frac{m^*}{m_{0}}\frac{\sigma g^*}{2}B_{y}(x)]\psi(x)=0,
\label{simple}
\ee
where we use units of length $l_{B}=\sqrt{\hbar c/eB_{0}}=1$, energy
$\hbar \omega_{c}=\hbar e B_{0}/m^{*} c=1$, and $B_{0}$ is the magnetic field scaling unit.
In the out-of-plane magnetic barrier system, it is essential to include electrostatic potential $U(x)$ in order to compensate for unwanted step-like potential barriers coming from the vector potential $\vec{A}$\cite{ihm}. However, the vector potential does not appear in Eq. \eqref{simple} since it becomes zero at the 2DEG plane and, therefore, it is not essential to include electrostatic potential $U(x)$ in the in-plane magnetic barrier system. Notice that the Zeeman term in Eq. \eqref{simple} plays a role of an effective potential barrier for spin-up($\sigma=+1$) electrons, while it acts like an effective potential well for spin-down($\sigma=-1$) electrons. Hereinafter, a magnetic barrier should be interpreted as an effective potential well for a spin-down($\sigma=-1$) electron.

Based on the above Shr\"odinger equation, transmission probability is calculated by the standard transfer matrix method.
Transfer matrices for the magnetic barriers $T_b$ and for the well confined by the barriers $T_{w}$ are ;
\bea
T_{b} &=&
\left( \begin{array}{cc}
\cosh \kappa d & (1/\kappa)\sinh \kappa d  \\
\kappa \sinh \kappa d  & \cosh\kappa d
\end{array} \right),
\\
T_{w} &=&
\left( \begin{array}{cc}
\cos k d & (1/k)\sin k d  \\
-k \sin k d  & \cos k d
\end{array} \right),
\eea
where $\kappa=\sqrt{2(U(x)-E)+k_{y}^2+\frac{m^*}{m_{0}}\frac{\sigma g^*}{2}B}$ and $k=\sqrt{2(E-U(x))+k_{y}^2}$.
The transmission probability $T^{\sigma}(E, k_{y})$ is obtained from the transmission coefficient $\tau$ of
the wavefunction after tunneling through the magnetic barriers by using the transfer matrices.
\bea
\tau &=&\frac{2 i k(C_{12}C_{21}-C_{11}C_{22})}{(C_{21}-k^{2}C_{12})-i k (C_{11}+C_{22})},
\label{matrix}
\eea
where $C_{ij}$ are elements of the transfer matrix $C=T_{b}\cdot T_{w}\cdot T_{b}$ for double barriers.

In our numerical calculation, since the better spin filtering effect is expected in a material with large $g^{*}$ and $m^{*}$\cite{ihm, nammee1}, material parameters of HgCdTe are used as follows:
the effective mass $m^{*}=0.01m_{0}$, $g$-factor $g^{*}=100$, energy unit $E_{0}=2.32$ meV,
magnetic length $l_{B}=57.5$ nm, and the magnetic scaling unit $B_{0}=0.2$ T.

\begin{figure}
\includegraphics[width=8.0cm]{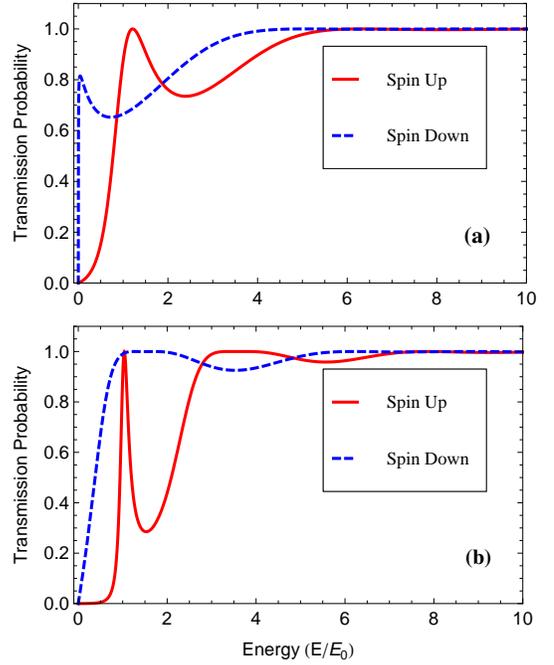}
\noindent \caption{ Transmission Probability $T^{\sigma}$ for the in-plane double magnetic barrier structure
as a function of incident electron's energy.
(a) $B=5$, $L=0.7$ and $d=0.7$, and
(b) $B=5$, $L=0.7$ and $d=1.4$  in dimensionless unit.}
 \label{Figure2}
\end{figure}
Figure 2 shows the transmission probability $T^{\sigma}(E,0)$  of the in-plane double magnetic barrier system with $U(x)=0$ for two different magnetic barrier widths.
For minimum energy requirement for electron tunneling, $k_{y}=0$ is chosen for qualitative calculation.
The transmission probability of both spins shows clear oscillating behavior as a function of
incident electron's energy, even when the electrostatic potential $U(x)$ is absent.

In Fig. 2(a), sharp resonance peaks are clearly seen for both spin-up and spin-down electrons.
As magnetic barrier width $d$ increases, however, the spin-down resonance peak is broadened due to the lack of barrier formation.
In Fig. 2(b), since the total structure length increases as the barrier width increases, energy difference between two resonant peaks for a spin-up electron decreases and the better negative spin polarization is achieved around $E/E_{0} \sim 1.5$ compared to the case of Fig. 2(a).

Figure 3 shows the transmission probability $T^{\sigma}(E,0)$ and corresponding
tunneling spin polarization $P(E, 0)$ for an in-plane triple magnetic barrier system as a function of incident electron's energy with electrostatic potential $U(x)=U_{0}\Theta(|x|-L/2) \Theta(L/2+d-|x|)$  applied.
\begin{figure}
\includegraphics[width=8.0cm]{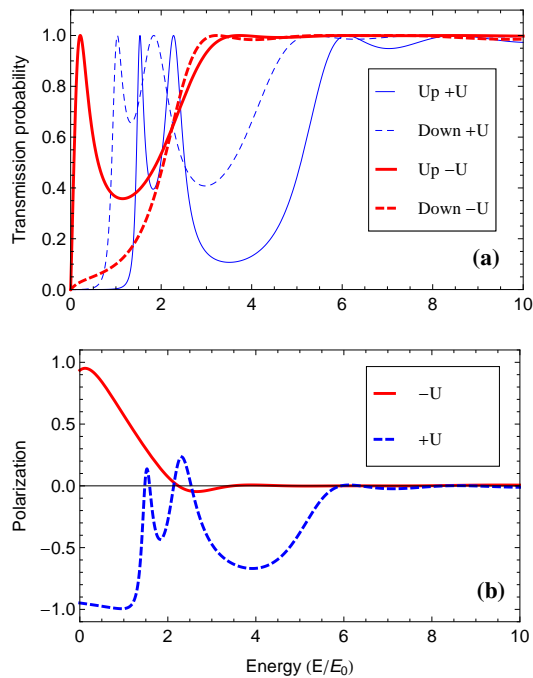}
\noindent \caption{(a) Transmission probability $T^{\sigma}$,
(b) corresponding tunneling spin polarization for a in-plane triple magnetic barrier structure with U(x)
as a function of incident electron's energy $E/E_{0}$.
Parameters, $B=2.5$, $L=0.7$, $d=0.7$, $k_{y}=0$ and $U_0=\pm 0.9B$ are used in the dimensionless unit.}
\label{Figure3}
\end{figure}
The tunneling spin polarization is defined by
\bea
P(E, k_{y})&=&\frac{T^{+} (E,k_{y})-T^{-} (E,k_{y})}{T^{+} (E,k_{y})+T^{-} (E,k_{y})}.
\label{pol}
\eea
As the number of barriers increases, a resonance peak splitting appears clearly.
In the triple case, $C_{ij}$ in Eq.\eqref{matrix} are obtained from $C=T_{b}\cdot T_{w}\cdot T_{b}\cdot T_{w}\cdot T_{b}$.
Since $U(x)$ affects on the height of effective  potential barrier, when positive electrostatic potential $U(x)$ is applied at the magnetic barrier region,
resonant peaks move to higher energy  than negative $U(x)$ case.
When $U_{0}> |\frac{m^*}{m_{0}}\frac{\sigma g^*}{2}B|$, the resultant effective potential for spin-down electrons becomes positive, and the spin-down electrons experience a barrier instead of a well. As a result, the resonance peak splitting appears also in the spin-down case, which is shown as the thin dashed line in Fig. 3(a).

Figure 3(b) shows tunneling spin polarization corresponding to Fig. 3(a).
Notice that in the low energy regime of $E/E_{0} \le 1.5$, the spin polarization can be switched between spin-up and spin-down by reversing the sign of the electric potential $U(x)$.
This implies that in-plane spin orientation of injected currents through the
triple magnetic barrier structure can be manipulated electrically.

In conclusion, the transmission properties of a two-dimensional electron gas system
with in-plane magnetic barriers are investigated.
Spin dependent resonance oscillation occurs in the transmission probability
even without electrostatic potential applied, although it can be used to control the spin current.
The transmission property and the current spin polarization can be manipulated efficiently  by the number of barriers as well as by electrostatic potential modulation.
As a result of this work, the in-plane triple magnetic barrier structure can serve as a good in-plane oriented spin selector which can be operated thoroughly by electrical modulation without any magnetic control.

\begin{acknowledgments}
We are grateful to J. W. Kim for helpful discussions.
This research was supported by Basic Science
Research Program through the National Research Foundation
of Korea(NRF) funded by the Ministry of Education,
Science and Technology (grant 2012R1A1A2006303
and 2010-0021328).
\end{acknowledgments}

\end{document}